\documentclass[11pt]{article}
\usepackage{amssymb, amsmath, multicol}

\newtheorem{prop}{Proposition}
\newtheorem{lem}{Lemma}
\newtheorem{thm}{Theorem}

\newtheorem{thmref}{Theorem}

\newcommand{\be}{\begin{equation}}
\newcommand{\ee}{\end{equation}}

\def\real{\mathop{\mathbb R}}      

\def\integ{\mathop{\mathbb Z}}      

\def\X{\mathop{\mathcal X}}        

\def\bx{{\bf x}}

\def\bv{{\bf v}}

\def\by{{\bf y}}

\begin{document}

\title{Denoising Deterministic Time Series}

\author{Steven P.\ Lalley and Andrew B.\ Nobel
\thanks{
S.P.\ Lalley is with the Department of Statistics, University of
Chicago, Chicago, IL  60637.  Email: lalley@galton.uchicago.edu \
A.B.\ Nobel is with the Department of Statistics, University of
North Carolina, Chapel Hill, NC  27599-3260. Email:
nobel@email.unc.edu \  His work supported in part by NSF Grant
DMS-9971964}}

\date{2002}

\maketitle

\begin{abstract}
This paper is concerned with the problem of recovering a
finite, deterministic time series from observations that are corrupted
by additive, independent noise.  A distinctive feature of this
problem is that the available data exhibit long-range dependence
and, as a consequence, existing statistical theory and
methods are not readily applicable.  This paper gives an
analysis of the denoising problem that extends recent
work of Lalley, but begins from first
principles.  Both positive and negative results are established.  The
positive results show that denoising is possible under somewhat
restrictive conditions on the additive noise.  The negative results
show that, under more general conditions on the noise, no
procedure can recover the underlying deterministic series.
\end{abstract}

\section{Introduction}

Recent interest in chaos has drawn the attention of statisticians
to deterministic phenomena that exhibit random behavior. While there is no
universally accepted definition of chaos, phenomena termed
``chaotic'' have generally been studied in the context of dynamical
systems, which provide mathematical models of physical
systems that evolve deterministically in time.
(Good introductions to dynamical systems and chaos for
non-specialists can be found in the texts of
Devaney \cite{Dev89} and Alligood {\it et al.} \cite{AlSaYo96}.)
In what follows we will consider a standard
model for dynamical systems, in which the
relevant states of the system form a compact subset $\Lambda$ of
$\real^d$.  The time evolution of the system is described by an
invertible map $F: \Lambda \to \Lambda$.
If at time $i$ the system is in state $x \in \Lambda$,
then at time $i+1$ it is in state $Fx$, and at time $i-1$
it is in state $F^{-1}x$.
That descriptions of this sort are, in a precise sense, generic
follows from Takens's embedding theorem \cite{Tak80,Aey81,SYC91}.
We do not assume that $F$ (or $F^{-1}$) is continuous.
Starting from an initial state $x \in \Lambda$
at time zero, the complete time evolution of the system is
described by the bi-infinite trajectory
\begin{equation*}
\begin{matrix}
\mbox{state} & \ldots & F^{-2}x  & F^{-1}x & x  & Fx & F^{2}x  & \ldots \\
\mbox{time} & \ldots & -2  & -1 & 0  & 1 &  2 & \ldots
\end{matrix}
\end{equation*}
Here $F^{i}$ is the
$i$-fold composition of $F$ with itself and $F^{-i}$ is the
$i$-fold composition of $F^{-1}$.
This model is deterministic: from exact
knowledge of the state of the system at any point in time, one may
reconstruct all the past and future states of the system
by repeated application of $F$ and $F^{-1}$.
To simplify notation in what follows, let $x_i = F^{i}x$,
$i \in \integ$, so that the initial state $x$ of the
system is denoted by $x_0$.

To date, most statistical analyses of dynamical systems
have been carried out in the context of dynamical
noise models.  In a dynamical noise model, the available
observations are assumed to be generated according to a
nonlinear autoregressive scheme of the form
$x_{i+1} = F x_i + \eta_i$, where $\{\eta_i\}$ are
independent, mean zero random vectors.  In this model,
random noise is ``folded'' into the dynamics at each step,
and the resulting sequence of states $x_i$ is not
purely deterministic.  In the presence of dynamical noise,
the observed states form
a discrete time, continuous state Markov Chain, and
estimating interesting features of the dynamics
(e.g.\ the map $F$) can often
be accomplished in part by an
appeal to traditional time series techniques.
Representative work can be found in references
\cite{WSSV85,FarSid87,Casd89,Casd92,KosYor90,NEGM92,LuSmi97,Tong90}.
An alternative approach to the map estimation problem
is described in \cite{Nob01}.

Of interest here is the so-called {\em observational noise model},
in which the available data are
observations (or measurements)
of an underlying deterministic system
that are corrupted by additive noise.
In this model our observations take the form
$y_i = x_i + \varepsilon_i$, where
$\{\varepsilon_i\}$ are independent, mean zero random vectors.
In contrast with the dynamical noise model,
the noise does not interact with the dynamics:
the deterministic character of the system, and
its long range dependence, are
preserved beneath the noise.
Due in part to this dependence,
estimation in the observational noise model
has not been broadly addressed by statisticians,
though the model captures important
features of many experimental situations.
Here we are interested in the problem of how
to recover the underlying time series $\{x_i\}$ from
the observations $\{y_i\}$.

\vskip.1in

\noindent
{\bf Denoising problem:}
Reconstruct the successive states $x_0,\ldots,x_n$ of the
deterministic system from observations of the form
\be
\label{obsn}
y_i \ = \ x_i + \varepsilon_i
    \ = \ F^i x_0 + \varepsilon_i
    \ \ \ 0 \leq i \leq n
\ee
where $\varepsilon_0, \ldots, \varepsilon_n \in \real^d$
are independent random vectors with mean zero.

\vskip.15in

Several versions of the denoising problem, and associated methods,
have previously been considered by a number of authors, including
Kostelich and Yorke \cite{KosYor90}, Davies \cite{Dav92}, Sauer
\cite{Sau92}, Kostelich and Schreiber \cite{KosSch93}. The methods
and results described here are motivated by recent work of
Lalley \cite{Lall99,Lall01}. MacEachern and Berliner
\cite{MacBer95} studied the problem of distinguishing trajectories
in the observational noise model when the noise distribution comes
from a suitable exponential family and established the asymptotic
normality of relevant likelihood ratios.

Though some features of denoising
can be found in more traditional statistical problems such as
errors in variables regression, deconvolution,
and measurement error modeling (c.f.\ \cite{CarRupSte95}),
other features distinguish it from these problems
and require new methods of analysis.
For example, in the denoising problem
the covariates $x_i$ are deterministically related (not i.i.d.\ or mixing),
the noise $\varepsilon_i$ is often bounded (not Gaussian),
and the noise distribution is usually unknown.

In the denoising problem the underlying states of the
observed deterministic system are of primary interest.
Denoising methods can also provide useful preprocessing
for other statistical analyses.
In the absence of noise, and under appropriate regularity conditions,
$x_0, x_1, \ldots$
can be used to estimate the map $F$ \cite{BosGue95,AdNob01,Nob01},
its invariant measure, entropy,
and Lyapunov exponents \cite{EKRC86}, or the fractal dimension
of its attractor (see \cite{Cut93}).
When observational noise is present,
consistent reconstructions $\hat{x}_0,\ldots,\hat{x}_n$ can
sometimes act as surrogates for the unobserved states in
estimation problems of this sort.
The surveys \cite{EckRu85,Ber92,Isham93,Jens93}
give an account of statistical problems in the study of
dynamical systems.  Formal limits to statistical inference
from dependent processes can be found in \cite{AdNob98,Adams97,Nob99}.
>From the viewpoint of statistical practice and theory, it is interesting
to ask whether estimation is still possible when noise removal is not,
but we will not address such issues here.

\section{Summary}

The next section contains several preliminary definitions and results that
will be used throughout the paper.  Section \ref{CD} describes two denoising
procedures.  The consistency of these procedures is established in
Theorems \ref{FXWN} and \ref{VWIN} under a boundedness assumption on the noise.
It is shown in Section \ref{NR} that, in a variety of settings,
consistent denoising is not possible when this assumption
is significantly relaxed.  Proofs of the positive (consistency)
results are given in Section \ref{PF12}; proofs of the negative results
are given in Section \ref{PFH}.

\section{Preliminaries}

Throughout this paper we assume that
$F: \Lambda \to \Lambda$ is an invertible map of a
compact set $\Lambda \subseteq \real^d$.
Of primary interest are maps that possess
an elementary form of sensitive dependence
on initial conditions.  Recall that $F$ is said to be expansive
if there exists $\Delta > 0$ such that
for every pair of vectors
$x,x' \in \Lambda$ with $x \neq x'$,
\[
\sup_{s \in \integ} |F^s x - F^s x'| \ > \ \Delta .
\]
The constant $\Delta$ is called a {\em separation
threshold} for $F$.
If $F$ is expansive then, beginning from
any two distinct initial states,
the corresponding bi-infinite trajectories of $F$ will,
at some (possibly negative) time $i$
be at least $\Delta$ apart.
Note that the separation
threshold $\Delta$ does not depend on $x$ or $x'$.

\vskip.1in

\noindent
{\bf Definition:} Let $F$ be an expansive map with separation
threshold $\Delta > 0$.  The separation time for $x \neq x'$ is
\[
s(x,x') = \min\{ |s| : |F^s x - F^s x'| > \Delta \} .
\]
For each $\alpha > 0$ define the {\em separation horizon}
\[
H(\alpha) \ = \ \sup\{ \, s(x,x') \ : \ |x - x'| \geq \alpha \, \} .
\]
Note that $\alpha \leq \alpha'$ implies $H(\alpha) \geq H(\alpha')$.
If $H(\alpha) < \infty$ for every $\alpha > 0$, then
then $F$ will be said to have {\em finite separation horizon}.

\vskip.1in

\begin{prop}
\label{FST}
If $F$ has finite separation horizon then the inverse function
\be
\label{ish}
H^{-1}(k) \, = \, \inf\{ \alpha > 0 : H(\alpha) \leq k \}
\ee
tends monotonically to zero as $k \to \infty$.
\end{prop}

\noindent
{\bf Proof:} The monotonicity of $H^{-1}$ follows from that of $H$.  If
$H^{-1}(k) \geq \alpha_0 > 0$ for every $k$, then $H(\alpha) = +\infty$
for $\alpha < \alpha_0$.

\vskip.15in

If $F: \Lambda \to \Lambda$ is invertible and continuous, then
$F^{-1}$ is continuous and $F$ is a homeomorphism
(see, {\em e.g.} \cite{Walt81}).
An elementary argument shows that
every expansive homeomorphism has finite separation horizon.

\begin{lem}
If $F: \Lambda \to \Lambda$ is an expansive homeomorphism, then
$F$ has finite separation horizon.
\end{lem}

\noindent
{\bf Proof:} Let $\Delta > 0$ be a separation threshold for $F$.  If
$H(\alpha) = +\infty$ for some $\alpha > 0$ then there exist pairs
of states $(x_n,x_n') \in \Lambda \times \Lambda$, $n \geq 1$, such that
$|x_n - x'_n| \geq \alpha$ for each $n$ and
$s(x_n,x'_n) \rightarrow \infty$.  As $\Lambda$ is compact, there exist
integers $n_1 < n_2 < \cdots$ and points $x,x' \in \Lambda$
such that $x_{n_k} \rightarrow x$ and $x'_{n_k} \rightarrow x'$.
Clearly $|x - x'| \geq \alpha$.  Moreover, as $F$ is continuous and
$s(x_{n_k},x'_{n_k}) \rightarrow \infty$, for each $m \geq 1$,
\[
\max_{|s| \leq m} |F^s x - F^s x'|
\ = \
\lim_{k \rightarrow \infty} \max_{|s| \leq m}
    |F^s x_{n_k} -F^s x'_{n_k}|
\ \leq \
\Delta .
\]
It follows that $H(\alpha) \geq s(x,x') = \infty$, which is a contradiction.

\subsection{Ergodic Transformations}

\noindent
{\bf Ergodic Transformation:} Let $\mu$ be a probability measure
on the Borel subsets of $\Lambda$.
A map $F: \Lambda \to \Lambda$ is said to preserve $\mu$ if
$\mu(F^{-1}B) = \mu(B)$ for each Borel set $B \subseteq \Lambda$.
A $\mu$-preserving map $F$ is said to be {\em ergodic} if
$F^{-1}B = B$ implies $\mu(B) \in \{0,1\}$, i.e.\ every $F$-invariant
set has $\mu$-measure zero or one.

\vskip.15in

The ergodic theorem generalizes the ordinary
law of large numbers and is an important tool
in understanding the asymptotic behavior of
dynamical systems.  It states that the time average
of a real-valued measurement along the trajectory
of an ergodic map $F$ will converge to the space
average of that measurement.

\begin{thmref}[Ergodic Theorem]
If $F:\Lambda \to \Lambda$ is $\mu$-preserving and ergodic,
and $f: \Lambda \to \real$ is such that
$\int|f| \, d\mu < \infty$, then
$n^{-1} \sum_{i=0}^{n-1} f(F^ix) \ \to \ \int f \, d\mu$
with probability one and in mean.
\end{thmref}

\section{Consistent Denoising}
\label{CD}

In this section we describe two consistent denoising methods
for deterministic time series, and provide a preliminary
analysis of their theoretical performance.

\subsection{Smoothing Algorithm D}

We first describe a denoising method originally
proposed by Lalley \cite{Lall99}, called
Smoothing Algorithm D.
Let the available
data be a sequence of vectors $y_0,\ldots,y_n$
defined as in (\ref{obsn}), and
let $k$ be a positive integer less than $\log n$.  For each
$l = k, \ldots, n-k$ define the index set
\be
\label{index}
A_n(l,k) \ = \ \{ \, j \, : \,  |y_{j+r} - y_{l+r}| \, \leq \, 3 \Delta / 5
                \mbox{ for } |r| \leq k \, \} .
\ee
Note that $l \in A_n(l,k)$ so that $A_n(l,k)$ is always non-empty.
For $l = k, \ldots, n-k$ define the denoising estimate
\be
\label{est1}
\hat{x}_{l,n} \ = \ \frac{1}{|A_n(l,k)|} \, \sum_{j \in A_n(l,k)} \, y_j
\ee
of $x_l$: set $\hat{x}_{l,n} = 0$ for other values of $l$.
To see how the
estimate is constructed, let $w(j,k) = (y_{j-k},\ldots,y_{j+k})$
contain the observations
in a window of length $2k+1$ centered at $y_j$.
The estimate $\hat{x}_{l,n}$ of $x_l$ is obtained by averaging all
those values $y_j$ for which $w(j,k)$ is close,
on a term by term basis, to $w(l,k)$.

\begin{thm}
\label{FXWN}
Let $F$ be an expansive map with
separation threshold $\Delta > 0$ and finite separation
horizon.  Suppose that $|\varepsilon_i| \leq \Delta / 5$ for
each $i \geq 0$.  If $k \to \infty$ and $k / \log n \to 0$ then
\[
\frac{1}{n-2k} \sum_{i=k}^{n-k} |\hat{x}_{i,n} - x_i| \ \to \ 0
\ \mbox{ as } \ n \to \infty
\]
with probability one for every initial vector $x \in \Lambda$.
\end{thm}

The in-probablity consistency of Smoothing Algorithm D
was first established in Theorem 1 of \cite{Lall99}
under the condition that $F$ is a $C^2$-diffeomorphism
and $\Lambda$ is a hyperbolic attractor (or the basin of
attraction of such a set).  A more general result for
expansive homeomorphisms is stated in
Theorem 2 of \cite{Lall01}.
Here these conditions are replaced by the weaker assumption of
finite separation horizon, and in-probability convergence is
strengthened to convergence with probability one.
The proof of Theorem \ref{FXWN} is given in Section \ref{PF12}.

\subsection{Implementation}

A naive implementation of smoothing algorithm D has running time
$O(n^2)$, where $n$ denotes the number of available observations.
More efficient, approximate, versions of the algorithm
with running time $O(n \log n)$ are investigated in \cite{Lall01}.
In simulations, Algorithm D and its approximations
have been used to successfully remove noise from trajectories of
the logistic map, the H\'enon attractor, and Smale's solenoid.
Informal studies have illustrated the failure of the algorithm
to remove uniform noise
whose support is comparable to the diameter of the
associated attractor.
These simulations lend empirical
support to Theorem \ref{FXWN} and the negative
results discussed below.

\subsection{Preliminary Analysis}
\label{PA}

Smoothing Algorithm D removes observation noise from the
trajectory of an expansive map by judicious averaging.
To understand why Theorem \ref{FXWN} holds,
fix $l$ between $k$ and $n-k$.
Together (\ref{obsn}) and (\ref{est1}) imply that
\be
\label{decomp1}
|x_l - \hat{x}_{l,n}| \ \leq \
\frac{\sum_{j \in A_n(l,k)} |x_l - x_j|}{|A_n(l,k)|} \ + \
\frac{| \sum_{j \in A_n(l,k)} \varepsilon_j \, |}{|A_n(l,k)|} .
\ee
The first term on the right hand side of (\ref{decomp1})
controls the bias of the estimate $\hat{x}_l$, and
the second controls its stochastic variation.  Regarding
the bias, note that
\begin{eqnarray*}
j \in A_n(l,k)
& \Rightarrow &
|y_{j+r} - y_{l+r}| \, \leq \, 3 \Delta / 5
  \ \mbox{ for } \ 1 \leq |r| \leq k \\
& \Rightarrow &
|x_{j+r} - x_{l+r}| \, \leq \, \Delta
  \ \mbox{ for } \ 1 \leq |r| \leq k \\
& \Rightarrow &
k \, \leq  \, H(|x_l - x_j|)
\ \Rightarrow \
|x_l - x_j| \, \leq \, H^{-1}(k) .
\end{eqnarray*}
Thus (\ref{decomp1}) implies that
\be
\label{xhinq1}
| x_l - \hat{x}_{l,n} |
\ \leq \
H^{-1}(k) \ + \
\frac{| \sum_{j \in A_n(l,k)} \varepsilon_j \, |}{|A_n(l,k)|} .
\ee
This yields the following bound
on the average denoising error:
\be
\label{averbnd}
\frac{1}{n-2k} \sum_{l=k}^{n-k} | x_l - \hat{x}_{l,n} |
\ \leq \
H^{-1}(k) \ + \
\frac{1}{n-2k} \sum_{l=k}^{n-k} \,
\frac{| \sum_{j \in A_n(l,k)} \varepsilon_j \, |}{|A_n(l,k)|} .
\ee
The upper bound $H^{-1}(k)$ on the average bias
depends on the map $F$ and the window width $k$,
but is independent of $n$ and $l$.
Moreover, $H^{-1}(k) \to 0$ by Proposition \ref{FST},
as $F$ has finite separation horizon and
$k \to \infty$.
Analysis of the stochastic variation is complicated by
the fact that the $\varepsilon_i$ are not independent when
summed over the {\em random} index set $A_n(l,k)$.
The details are given in the appendix (see in particular
inequality (\ref{inq2}) and Lemma \ref{comb}).

The analysis above suggests a more adaptive
version of Smoothing
Algorithm D that offers improved performance under
somewhat stronger conditions.
Fix $l$ for the moment and consider inequality
(\ref{xhinq1}).  It can be seen that the window width $k$
plays a role analogous to inverse bandwidth in kernel type
estimators.  Monotonicity of $H^{-1}$ ensures that
the bias of $\hat{x}_{l,n}$ decreases as $k$ increases.
On the other hand, as $k$ increases, the index set
$A_n(l,k)$ gets smaller,
and the variability of the estimate will increase
as one averages over fewer noise variables $\varepsilon_j$.
One modification of Smoothing Algorithm D, analogous to
local bandwidth selection, is to
adaptively select a window width for each location $l$.
This is considered in more
detail below.

\subsection{Denoising with a Variable Length Window}
\label{VLW}

Here new denoising estimates $\tilde{x}_{l,n}$ are described.
Let the index sets $A_n(l,k)$ be defined as in (\ref{index}).
The new
estimates are based on windows whose widths are chosen adaptively
to ensure that $|A_n(l,k)|$ is sufficiently large.
For $l = \log n, \ldots, n - \log n$ define
\be
\label{aww}
k_{l,n} \ = \ \max\{ 1 \leq k \leq \log n \, : \, |A_n(l,k)| \geq n / \log n \},
\ee
and set $k_{l,n} = 0$ if $|A_n(l,1)| < n / \log n$.
For the same values of $l$, define denoising estimates
\be
\label{estn2}
\tilde{x}_{l,n} \ = \ \frac{\sum_{j \in A_n(l,k_{l,n})} y_j}{|A_n(l,k_{l,n})|} .
\ee
Set $\tilde{x}_{l,n} = 0$ if $k_{l,n} = 0$.
Strong consistency of the estimates $\tilde{x}_{l,n}$
requires that the trajectory under study exhibit
a natural recurrence property.

\vskip.2in

\noindent
{\bf Definition:} A point
$x \in \Lambda$ with trajectory $x_i = F^ix$ will be called
strongly recurrent if there is a finite cover
${\cal O}$ of $\Lambda$ such that (i)
every $O \in {\cal O}$ has diameter less than $\Delta / 5$,
and (ii) for each $r \geq 1$
and each choice of sets $O_1,\ldots,O_r \in {\cal O}$ either
\be
\label{apdens1}
\sum_{i=0}^\infty
I\{ x_i \in O_1, \ldots, x_{i+r-1} \in O_r \}
\ < \ \infty
\ee
or
\be
\label{apdens2}
\liminf_{n \to \infty}
\frac{1}{n} \sum_{i=0}^{n-1}
I\{ x_i \in O_1, \ldots, x_{i+r-1} \in O_r \}
\ > \ 0 .
\ee
Conditions (\ref{apdens1}) and (\ref{apdens2}) ensure that
if the forward trajectory of $F$ starting from $x$
visits the product set $O_1 \times \cdots \times O_r$ infinitely
often, then it does so a non-negligible fraction of the time.

\vskip.2in

Recall that $F$ is said to preserve a probability measure $\mu$
on the Borel subsets of $\Lambda$ if
$\mu(F^{-1}B) = \mu(B)$ for each Borel set $B \subseteq \Lambda$,
and that
$\mu$-preserving map $F$ is said to be ergodic if
$F^{-1}B = B$ implies $\mu(B) \in \{0,1\}$, i.e.\ every $F$-invariant
set has $\mu$-measure zero or one.
Strongly recurrent points are the norm in measure preserving
systems.

\begin{prop}
If $F$ preserves a measure $\nu$ on $\Lambda$ and is ergodic then
$\nu$-almost every $x \in \Lambda$ is strongly recurrent.
\end{prop}

\noindent
{\bf Proof:}
Let ${\cal O}$ be any finite open cover of $\Lambda$ by sets
having diameter less than $\Delta / 5$.  Fix sets
$O_1,\ldots,O_r \in {\cal O}$.  Note that
$x_i \in O_1,\ldots,x_{i+r-1} \in O_r$ if and only if
$x_i =  F^ix \in O'$ where $O' = \cap_{j=1}^r F^{-j+1} O_{j}$.
If $\nu(O') > 0$, the ergodic theorem ensures that
\[
\lim_{n \to \infty}
\frac{1}{n} \sum_{i=0}^{n-1}
I\{x_i \in O_1,\ldots,x_{i+r-1} \in O_r\}
\ = \
\lim_{n \to \infty}
\frac{1}{n} \sum_{i=0}^{n-1}
I\{ F^i x \in O' \}
\ = \ \nu(O') > 0
\]
with $\nu$-probability one.
On the other hand, if $\nu(O') = 0$ then
$\sum_{i=1}^\infty \nu(F^{-i}O') = 0$ and consequently
$\nu\{ F^i x \in O' \mbox{ infinitely often} \} = 0$ by
the first Borel Cantelli lemma.

\begin{thm}
\label{VWIN}
Let $F$ be an expansive map with separation threshold $\Delta > 0$
and finite separation horizon.  If
$|\varepsilon_i| \leq \Delta / 5$ for each $i \geq 0$, then
for every strongly recurrent initial vector $x \in X$,
\[
\max\{ \, |\tilde{x}_{l,n} - x_l| \, : \, \log n \leq l \leq n - \log n \, \}
\ \to \ 0
\]
with probability one as $n$ tends to infinity.
\end{thm}

Performance bounds of this sort for Smoothing Algorithm D
are established in \cite{Lall99} under the stronger assumption that
$F$ is a $C^2$-diffeomorphism and that
$\Lambda$ is an Axiom A basic set.

\section{Negative Results}
\label{NR}
One distinctive (and restrictive)
feature of Theorems \ref{FXWN} and \ref{VWIN} is the assumption that
the noise $\varepsilon_i$ is bounded in absolute value by a
fraction of the separation threshold $\Delta$.
In light of the popularity and widespread study of Gaussian noise,
it is natural to ask if denoising is possible
when the $\varepsilon_i$ are normally distributed,
perhaps under some constraints on the component-wise variances.
Surprisingly, the answer is often ''no''.
Lalley \cite{Lall99} shows that for many smooth dynamical systems
no scheme can successfully remove Gaussian noise,
even in the weak sense of Theorem \ref{FXWN}.
In this section we extend and generalize this result.
Our proof covers the Gaussian case, generalizations of the Gaussian
case to noise distributions supported on
all of $\real^d$ (stated in \cite{Lall01}),
and the case of noise distributions with bounded support.

Suppose, as in the previous section, that
$\{ x_i = F^ix : i \in \integ \}$ is the trajectory of a
fixed initial vector $x \in \Lambda$, and that observations of
$x_i$ are subject to additive noise, and can be modeled as
random vectors
\be
\label{ydef}
y_i \ = \ x_i + \varepsilon_i \ \ \ i \in \integ
\ee
where $\ldots, \varepsilon_{-1}, \, \varepsilon_0, \,
\varepsilon_1, \ldots \in \real^d$ are independent, mean-zero
random vectors having a common distribution $\eta$ on $\real^d$.
We assume in what follows that the $\varepsilon_i$ are defined
on a common underlying probability space $(\Omega, {\cal F}, P)$.
Of interest here are several related problems, which
may be informally expressed as follows.

\begin{quote}
{\bf Problem 1:} Identify the initial state $x \in \Lambda$ from
observation of the infinite sequence $\{y_i : i \in \integ\}$.
\end{quote}

\begin{quote}
{\bf Problem 2:}
Consistently identify the initial state $x \in \Lambda$ from
observations $y_{-n},\ldots,y_{n}$, in the limit as $n \to \infty$.
\end{quote}

\begin{quote}
{\bf Problem 3:}
Estimate the states $x_1,\ldots,x_n \in \Lambda$ from
observation of $y_1,\ldots,y_n$.
\end{quote}

\noindent
It is evident that Problem 1 is easier than Problem 2, as in the former
we have access to all the available data at the outset.
It is also clear that an answer to Problem 2 might be used, in
conjunction with shifts of the observations, to answer Problem 3.
Problem 3 is just the denoising problem considered in the previous
section.

It is shown in Theorem \ref{HOM} below that for
distinguished states $x$
and noise distributions $\eta$, neither
Problem 1 nor Problem 2 has a solution.  This negative
result is then used to establish Theorem \ref{dns-cex},
which states that, for suitable dynamical
maps $F$ and noise distributions $\eta$,
consistent denoising is impossible.

\subsection{Distributional Assumptons}

The negative results in Theorems \ref{HOM} and \ref{dns-cex}
require that the distribution $\eta$ of
$\varepsilon_i$'s be smooth and
has sufficiently large support.  Here we give a precise
statement of these conditions.

Suppose first that $\eta$ is absolutely continuous, having a density
$f$ with respect to $d$-dimensional Lebesgue measure $\lambda$.
Recall that if $A$ is a Borel subset of $\real^d$,
$u \in \real^d$ is any vector and $r > 0$, then
\[
A + u \ = \ \{ v + u : v \in A \}
\ \ \mbox{ and } \ \
A^r \ = \ \{ u \, : \, |u - v| < r \mbox{ for some } v \in A \}
\]
are also Borel subsets of $\real^d$.  For $v \in \real^d$ and $r > 0$, let
$B(v,r) = \{ u : |u - v| < r \}$
be the Euclidean ball of radius $r$ centered at $v$.
Let $S = \{ v : f(v) > 0 \}$ be the support of the density $f$ of $\eta$.
Let $\overline{S}$ and $S^o$ denote the closure and interior of $S$,
respectively, and let
$\partial S = \overline{S} \setminus S^o$ be its boundary.
Finally, let $\rho = \max\{ |u-v| : u, v \in \Lambda \}$ be the diameter
of $\Lambda$.  Note that $\rho$ is finite as $\Lambda$ is compact.
We make the following assumptions concerning $\eta$:

\be
\label{cond1}
\limsup_{|z| \to 0} \frac{1}{|z|} \,
\int_{S \cap (S-z)} \left| \, \log \frac{f(w+z)}{f(w)} \right|
     f(w) \, dw
\ < \ \infty ,
\ee

\be
\label{cond2}
\limsup_{r \searrow 0} \frac{1}{r} \,
\eta( (\partial S)^r )
\ < \ \infty , \mbox{ and }
\ee

\be
\label{cond3}
S \supseteq \, B(0, 3 \rho /2).
\ee

\noindent
Assumption (\ref{cond1}) states that $\log f$ is Lipschitz
continuous on the average.
Assumption (\ref{cond2}) says that the measure of those
points within distance
$r$ of $\partial S$ decreases at least linearly with $r$.
Assumption (\ref{cond3}) states that $S$ contains a sphere whose radius
is significantly larger than the diameter $\rho$ of $\Lambda$.
It is enough that assumptions
(\ref{cond2}) and (\ref{cond3}) hold for {\em some} version $f$ of
$d\eta/d\lambda$.
Note that (\ref{cond2}) and (\ref{cond3}) are trivially
satisfied if $S = \real^d$.

\vskip.3in

\noindent
{\bf Example 1:} If $\eta$ is multivariate Gaussian and has a
covariance matrix of full rank, then
assumptions (\ref{cond2}) and (\ref{cond3}) are immediate, and
one may readily verify that assumption (\ref{cond1}) holds.

\vskip.2in

\noindent
{\bf Example 2:} Suppose that $\eta$ has a density $f$ with compact
support $S$ satisfying (\ref{cond3}),
and suppose further that $f$ is Lipschitz continuous on $S$.
Then $f$ is bounded away from zero and infinity on $S$ and
one may verify that (\ref{cond1}) holds.  Satisfaction of
(\ref{cond2}) requires, in addition, that the boundary of $S$
be regular.  To quantify this, let
$N(\partial S,r)$ denote the least number of Euclidean
balls of radius $r > 0$ needed to cover $\partial S$.
If $N(\partial S, r) \leq c \, (1/r)^{d-1}$ for some
$c < \infty$ and each $0 < r < r_0$, then
\[
\eta((\partial S)^r)
\ \leq \
c' \, \sup_{x \in S} |f(x)| \,
N(\partial S, r) \cdot r^d
\ \leq \
c' c \, \sup_{x \in S} |f(x)| \cdot r
\]
for a suitable normalizing constant $c'$,
and (\ref{cond2}) follows.
The bound $N(\partial S, r) \leq c \, (1/r)^{d-1}$ implies,
in particular, that the
box counting dimension of $\partial S$ is $d-1$.
Assumption (\ref{cond2}) is satisfied, for example,
if $\eta$ is the uniform distribution on
$B(0, 3 \rho /2)$, or the uniform distribution
on a cube of side length $3 \rho /2$
centered at the origin.

\subsection{Homoclinic Pairs}

Let $x$ and $x'$ be distinct initial states in $\Lambda$,
with corresponding
trajectories $\{ x_i = F^i x : i \in \integ \}$ and
$\{ x_i' = F^i x' : i \in \integ \}$.
Suppose that we wish to distinguish
$x$ and $x'$ on the basis of their trajectories.
In the absence of noise, and with knowledge of $F$,
this is always possible:
from observation of any $x_i$ one can recover
$x$, and from observation of any $x_j'$ one can
recover $x'$.  However, when observation noise is present, this simple
inversion process is no longer applicable.
Recall that $y_i = x_i + \varepsilon_i$, $i \in \integ$,
are noisy observations of the trajectory of $x$.  Let
\be
\label{y'def}
y_i' \ = \ x_i' + \varepsilon_i \ \ \ i \in \integ
\ee
be observations of the trajectory of $x'$, corrupted by the
same additive noise sequence.
Define $\X$ to be the set of all bi-infinite sequences
$\bv = \ldots, v_{-1}, v_0, v_1, \ldots$ with
$v_i \in \real^d$, and let ${\cal S}$ be the product
sigma field for $\X$
generated by the finite dimensional Borel cylinder sets.
For fixed $x, x'$ the sequences
\[
\by \ = \ (\ldots, y_{-1}, y_0, y_1, \ldots)
\ \ \mbox{ and } \ \
\by' \ = \ (\ldots, y_{-1}', y_0', y_1', \ldots)
\]
are random elements of $(\X,{\cal S})$, defined on the
underlying probability space
$(\Omega, {\cal F}, P)$.  Consider the following
special case of Problem 1 above.

\begin{quote}
{\bf Question 1:} Is there a measurable function
$\phi: \X \to \real^d$ such that
$\phi(\by) = x$ and
$\phi(\by') = x'$
with probability one?
\end{quote}

\noindent
Intuitively, it will be more difficult to identify $x$ and $x'$ in
the presence of noise
if their trajectories stay close to each other across
time.  The notion of a strongly homoclinic pair is one
way of making this precise.

\vskip.1in

\noindent
{\bf Definition:}
A pair $(x,x')$ of distinct states in $\Lambda$ is said to be
{\em strongly homoclinic} for $F$ if their bi-infinite
trajectories are such that
\be
\label{st-hom}
\sum_{i \in \integ} |F^i x - F^i x'| \ < \ \infty
\ee

\vskip.1in

As noted in \cite{Lall99}, homoclinic pairs exist and
are common in many smooth dynamical systems.  It is worth
noting that
the existence of a separation threshold does not preclude
the existence of homoclinic
pairs, as the separation of $F^i x$ and $F^i x'$ need only occur
for {\em one} value of $i$.  Theorem \ref{homoclin} below shows that
the answer to Question 1 is
"no" when $x$ and $x'$ are strongly homoclinic.
The proof is given in Section \ref{PFH}.

\begin{thm}
\label{HOM}
Suppose that the distribution $\eta$ of $\varepsilon_i$ satisfies
conditions (\ref{cond1})--(\ref{cond3}).  If $x$ and $x'$ are strongly
homoclinic, then for every measurable function
$\phi : \X \to \real^d$,
\[
E[ \, | \phi(\by) - x \, | \, + \,
      | \phi(\by') - x' \, | \, ]
\ > \ 0 .
\]
\end{thm}

\noindent
{\bf Remark:} Among the functions $\phi$ included in the theorem are
those that incorporate knowledge of the dynamical map and the two
possible initial states. Thus even with
knowledge of $\{x, x'\}$ and $F$, and even
with access to the entire noisy trajectory, one cannot recover
the initial state of the system with certainty.

\subsection{Negative Results for Denoising}

Suppose now that $F:\Lambda \to \Lambda$ preserves a Borel measure $\mu$ on
$\Lambda$ and is ergodic.  Let $X \sim \mu$ be independent of
$\{ \varepsilon_i \}$ and define
\be
\label{UYt}
X_i \ = \ F^i X , \ \ \
Y_i \ = \ X_i + \varepsilon_i
\ \ \ \ i \in \integ
\ee
where the $\varepsilon_i$
are i.i.d.\ with distribution $\eta$.
Then $\{ (X_i,Y_i) : i \in \integ \}$
is a stationary ergodic process taking values in
$\real^d \times \real^d$.  Our principal negative
result applies to dynamical systems that admit a
homoclinic coupling.

\vskip.2in

\noindent
{\bf Definition:} A $\mu$ preserving transformation $F: \Lambda \to \Lambda$
admits a {\em homoclinic coupling} if
on some probability space one may define random vectors
$X$ and $X'$ such that
\begin{enumerate}
\item $X$ and $X'$ take values in $\Lambda$
\item $X$ and $X'$ have distribution $\mu$
\item $(X,X')$ is strongly homoclinic for $F$ with positive probability.
\end{enumerate}

\vskip.2in

For systems admitting a homoclinic coupling, strongly homoclinic
pairs are relatively common.
When a homoclinic coupling exists we may ensure, by means
of a standard product construction, that the pair $(X,X')$
is defined on the same probability space
as, and is independent of, the noise variables $\varepsilon_i$.
It is shown in
\cite{Lall99} that many common models of smooth dynamical
systems, for example uniformly hyperbolic (and Axiom A)
$C^2$-diffeomorphisms, admit homoclinic couplings.

\vskip.15in

\noindent
{\bf Definition:} A denoising procedure is a collection of
measurable maps $\psi_{n,i}: (\real^d)^n \to \real^d$, with
$n \geq 1$, and $i = 1,\ldots,n$.  The
procedure $\{ \psi_{n,i} \}$ is weakly consistent for
a process $\{ (X_i,Y_i) \}$ if
\[
E\left[ \frac{1}{n} \sum_{i=1}^n
|\psi_{n,i}(Y_1,\ldots,Y_n) - X_i| \right]
\ \to \ 0
\]
as $n$ tends to infinity.

\vskip.2in

\begin{thm}
\label{dns-cex}
Suppose that $F : \Lambda \to \Lambda$ is a $\mu$-preserving
ergodic transformation
that admits a homoclinic coupling $(X,X')$.
If the distribution $\eta$ of $\varepsilon_i$ satisfies
conditions (\ref{cond1}) - (\ref{cond3}) then no denoising procedure
is weakly consistent for the process $\{ (X_i,Y_i) \}$ defined in (\ref{UYt}).
\end{thm}

\noindent
{\bf Proof:}
Assume, without loss of generality, that $X$ is the first component
of a homoclinic coupling $(X,X')$ for $F$.
Let $X_i' = F^i X'$ and $Y_i' = X_i' + \varepsilon_i$ for
$i \in \integ$.
Fix a denoising scheme $\{ \psi_{n,i} \}$ and
assume by way of contradiction that
\be
\label{temp}
E\left[ \frac{1}{n} \sum_{i=1}^n
|\psi_{n,i}(Y_1,\ldots,Y_n) - X_i| \right]
\ \to \ 0.
\ee
The joint distribution of
$\{(X_i,Y_i)\}$ is the same as that of
$\{(X_i',Y_i')\}$ and, therefore (\ref{temp}) implies that
\be
\label{temp'}
E\left[ \frac{1}{n} \sum_{i=1}^n
|\psi_{n,i}(Y_1',\ldots,Y_n') - X_i'| \right]
\ \to \ 0.
\ee
For each $n \geq 1$ define
\[
\phi_n(v_{-n},\ldots,v_n) \ = \
\frac{1}{n} \sum_{i=1}^n \psi_{n,i}(v_{1-i},\ldots,v_{n-i}) .
\]
The stationarity of $\{(X_i,Y_i)\}$ implies that
\begin{eqnarray*}
E| \, \phi_n(Y_{-n},\ldots,Y_n) \, - \, X \, |
& = &
E\left| \frac{1}{n}
    \sum_{i=1}^n \psi_{n,i}(Y_{1-i},\ldots,Y_{n-i})
        \, - \, X \, \right| \\
& \leq &
\frac{1}{n} \sum_{i=1}^n
   E| \, \psi_{n,i}(Y_{1-i},\ldots,Y_{n-i}) \, - \, X \, | \\
& = &
\frac{1}{n} \sum_{i=1}^n
   E| \, \psi_{n,i}(Y_1,\ldots,Y_n) \, - \, X_i \, | ,
\end{eqnarray*}
which tends to zero by (\ref{temp}).  An analogous argument using
(\ref{temp'}) shows that
\[
E| \, \phi_n(Y_{-n}',\ldots,Y_n') \, - \, X' \, | \ \to \ 0 .
\]
If $H$ is the event that $(X,X')$ is strongly homoclinic for $F$
then, letting $v_i^j = v_i,\ldots,v_j$,
\begin{eqnarray*}
0
& = &
\lim_{n \to \infty} E \left[
   | \phi_n(Y_{-n}^n) \, - \, X | \, + \,
   | \phi_n(Y_{-n}^{'n}) \, - \, X' | \right] \\
& \geq &
\liminf_{n \to \infty}
E \left[
( \, | \phi_n(Y_{-n}^n) \, - \, X | \, + \,
   | \phi_n(Y_{-n}^{'n}) \, - \, X' | \, ) \cdot I_H \right] .
\end{eqnarray*}
It follows from Theorem \ref{HOM}
and the assumption that $P(H) > 0$ that
the last term above is positive.
As this leads to an evident contradiction,
(\ref{temp}) cannot hold, and the proof is complete.

\subsection{Some Refinements}

The proof of Theorem \ref{dns-cex} shows that
the values of $X_1,X_2,\ldots$ are not estimable, even
if one is given access to the
{\em entire sequence}
$\ldots, Y_{-1}, Y_0, Y_1, \ldots$
generated by $X$ and the noise.
In particular, there is no function $\psi: \X \to \real^d$
such that
\[
E\left[ \frac{1}{n} \sum_{i=1}^n
|\, \psi(\ldots, Y_{i-1}, Y_i, Y_{i+1}, \ldots)
    \, - \, X_i \, | \right]
\ \to \ 0
\]
as $n$ tends to infinity.

Another question that arises is how Theorem \ref{dns-cex} bears
on the problem of
denoising a trajectory arising from a fixed (non-random) initial
vector $x \in \Lambda$.  It follows immediately that if $x_i = F^i x$ and
$y_i = x_i + \varepsilon_i$, then
there is no denoising procedure
such that
\[
E\left[ \frac{1}{n} \sum_{i=1}^n
|\, \psi_{n,i}(y_1,\ldots,y_n) \, - \, x_i \, | \right]
\ \to \ 0
\]
for $\mu$-almost every initial state $x \in \Lambda$.
For denoising procedures satisfying a natural fading-memory
property, this conclusion may be strengthened.  Let us say
that a procedure $\{ \psi_{n,i} \}$ has fading memory if,
with $Y_i$ defined as in (\ref{UYt}), for each $k \geq 1$,
\[
\lim_{n \to \infty}
E\left[ \frac{1}{n-k} \sum_{i=k+1}^n
|\, \psi_{n-k,i-k}(Y_{k+1},\ldots,Y_n) \, - \,
    \psi_{n,i}(Y_{1},\ldots,Y_n) \, | \right]
\ = \ 0
\]
Averaging methods such as Smoothing
Algorithm D posess the fading memory property.
Under the conditions of Theorem \ref{dns-cex},
it can be shown that if
$\{ \psi_{n,i} \}$ has fading memory, then
\[
\limsup_{n \to \infty}
E\left[ \frac{1}{n} \sum_{i=1}^n
|\, \psi_{n,i}(y_1,\ldots,y_n) \, - \, x_i \, | \right]
\ > \ 0
\]
for $\mu$-almost every initial state $x \in \Lambda$.  Thus
successful denoising is not possible starting from almost
any initial state.

\section{Proof of Theorems \protect{\ref{FXWN}} and \protect{\ref{VWIN}}}
\label{PF12}

\subsection{McDiarmid's Inequality}
\label{MCD}

McDiarmid's inequality is a special case of what is known as the concentration of
measure phenomena.  The basic idea is the following.
If $f(x_1,\ldots,x_n)$ be a function that does not depend
too strongly on the value of any single argument, and
if $X_1,\ldots,X_n$ are independent random variables, then
$f(X_1,\ldots,X_n)$ will be close to
$Ef(X_1,\ldots,X_n)$ with high probability.  Put another way,
the distribution of $f(X_1^n)$ will be {\em concentrated} around its mean.
For a proof and discussion of the following result,
see \cite{Mcd89,DeGyLu96}.

\begin{thmref}[McDiarmid]
\label{McDiarmid}
Let $X_1,\ldots,X_n$ be independent random variables taking
values in a set $A \subseteq \real$ and let $f : A^n \rightarrow \real$.
For $i=1, \ldots,n$ define
\be
\label{coeff}
v_i \ = \ \sup | f(x_1^n) - f(x_1^{i-1}, x_i', x_{i+1}^n) | ,
\ee
where the supremum is over all numbers $x_1, \ldots, x_n, x_i' \in A$.
Then for every $t>0$
\be
P\{ |f(X_1^n) - E f(X_1^n)| > t \}
\ \leq \
2 \, \exp \left\{ \frac{-2 t^2}{\sum_{i=1}^n v_i^2} \right\} .
\ee
\end{thmref}

\subsection{Analysis of Stochastic Variability}

Here we derive exponential inequalities for the final term
in (\ref{xhinq1}), which governs the stochastic variability of
the estimate $\hat{x}_{l,n}$.
Define $U_n(l,k) = \sum_{j \in A_n(l,k)} \varepsilon_j$.

\begin{lem}
\label{case} If $H(\Delta/5) \leq k < n/2$ and $k \leq l \leq n-k$
then
\[
U_n(l,k) \ = \
\sum_{j=k}^{n-k} \varepsilon_j \, I\{|x_l-x_j| \leq \Delta / 5 \}
\prod_{1 \leq |s| \leq k} I\{|y_{l+s} - y_{j+s}| \leq 3 \Delta / 5 \}
\]
\end{lem}

\noindent
{\bf Proof:} Note that $U_n(l,k)$ can be written in the form
\[
U_n(l,k)
\ = \
\sum_{j=k}^{n-k} \varepsilon_j \prod_{|s| \leq k}
    I\{|y_{l+s} - y_{j+s}| \leq 3 \Delta / 5 \}
\]
Fix $j$ and define the quantities
\[
W_0 \ = \ \prod_{|s| \leq k}
          I\{|y_{l+s} - y_{j+s}| \leq 3 \Delta / 5 \}
\]
and
\[
W_1 \ = \ I\{|x_l - x_j| \leq \Delta / 5 \}
\prod_{1 \leq |s| \leq k} I\{|y_{l+s} - y_{j+s}| \leq 3 \Delta / 5 \} .
\]
It suffices to show that $W_0 = W_1$.  If $|x_l-x_j| \leq \Delta / 5$
then $|y_l - y_j| \leq 3 \Delta / 5$ and the desired equality is immediate.
Suppose then that $|x_l-x_j| > \Delta / 5$, in which case $W_1 =0$.
If in addition $W_0 = 1$, then $|x_{l+s} - x_{j+s}| \leq \Delta$
for $|s| \leq k$, which implies that
$|x_l - x_j| \leq H^{-1}(k) \leq \Delta / 5$.
As this is a contradiction,
$W_0$ must be zero, and the lemma is established.

\vskip.2in

\begin{lem}
\label{expineq}
Let $L = \Delta / 5$ be an upper bound on $|\varepsilon_i|$.
Fix $n \geq 1$ and integers
$l,k$ satisfying the conditions of Lemma \ref{case}.
Then for every $t > 0$,
\[
P \{ |U_n(l,k)| > t \}
\ \leq \
2 \exp \left\{ \frac{-2t^2}{nL^2(2k+1)^2} + \frac{4t}{nL(2k+1)} \right\} ,
\]
and in particular
\[
P \{ |U_n(l,k)| > t \}
\ \leq \
2 \exp \left\{ \frac{-t^2}{2nL^2(2k+1)^2} \right\}
\]
for $t \geq 2L(2k+1)$.
\end{lem}

\noindent
{\bf Proof:}
Define $\tilde{U}$ by excluding indices $j=l-k,\ldots,l+k$
from the sum defining $U_n(l,k)$, more precisely
\[
\tilde{U}
\ = \
\left(\sum_{j=k}^{l-k-1} + \sum_{j=l+k+1}^{n-k} \right)
\varepsilon_j \, I\{ |x_l - x_j| \leq \Delta / 5 \}
\prod_{1 \leq |s| \leq k}
I\{ |y_{l+s} - y_{j+s}| \leq 3 \Delta / 5 \} ,
\]
with the understanding that the first sum is zero
if $l \leq 2k$, and the second sum is zero if
$l \geq n - 2k$.
Then $|U_n(l,k) - \tilde{U}| \leq (2k+1) L$, and
as $\varepsilon_j$ is independent of the other products in
the $j$'th summand, $E \tilde{U} = 0$.
Suppose for the moment that the values of
$\varepsilon_{l-k}, \ldots, \varepsilon_{l+k}$ have been fixed.
In this case $y_{l-k},\ldots, y_{l+k}$ are fixed, and
$\tilde{U}$ is a function of $n - (2k+1)$ independent random
variables
$\Theta = \{ \varepsilon_j : j = 1,\ldots,l-k-1,l+k+1,\ldots,n \}$.
Let $f$ be such that $\tilde{U} = f(\Theta)$.
Changing any $\varepsilon_j \in \Theta$ will change $y_j$, and can
affect at most $2k+1$ terms in the sum defining
$\tilde{U}$; thus the coefficient $v_j$ defined in (\ref{coeff}) is
at most $(2k+1)L$.  As
$E( \tilde{U} \, | \, \varepsilon_{l-k}^{l+k} ) = E \tilde{U} = 0$,
McDiarmid's inequality implies that
\begin{eqnarray*}
P( |\tilde{U}| >t \ | \ \varepsilon_{l-k}^{l+k} )
& \leq &
2 \exp \left\{ \frac{-2t^2}
{\left(\sum_{j=k}^{l-k-1}+\sum_{j=l+k+1}^{n-k} \right) ((2k+1)L)^2} \right \} \\
& \leq &
2 \exp \left\{ \frac{-2t^2}{n \, (2k+1)^2 \, L^2} \right \} .
\end{eqnarray*}
Taking expectations, the same inequality holds for $P \{ |\tilde{U}| > t\}$.
The first of the stated inequalities follows from the fact that
$|U_n(l,k) - \tilde{U}| \leq (2k+1) L$, and the second follows from the
first by a straightforward calculation.

\vskip.15in

\noindent
{\bf Definition:} Let
$V_n(l,k) = |A_n(l,k)|^{-1} \sum_{j \in A_n(l,k)} \varepsilon_j$
be the stochastic term appearing in (\ref{xhinq1}).
For each $m \geq 1$ and $1 \leq k \leq n/2$ define
\[
L_n(m,k) = \{ l : | A_n(l,k) | \geq m \mbox{ and } k \leq l \leq n - k \}
\]
to be the set of indices $l$ for which at least $m$ length-$k$ matches
are found.

\vskip.1in

As an immediate corollary of Lemma \ref{expineq} we may derive
bounds on the probability that one of the terms $V_n(l,k)$ with
$|A_n(j,k)| \geq m$ exceeds a given constant $\delta > 0$.
In particular, treating a maximum over the empty set as zero,
we find that
\begin{eqnarray}
P \left\{ \max_{l \in L_n(m,k)} |V_n(l,k)| > \delta \right\}
& = &
P \left\{ \max_{l \in L_n(m,k)}
          \frac{|U_n(l,k)|}{| A_n(l,k)|} > \delta \right\}
          \nonumber \\[.05in]
& \leq &
P \left\{ \max_{l \in L_n(m,k)} |U_n(l,k)| > \delta m \right\}
          \nonumber \\[.07in]
& \leq &
n \cdot \max_{l} P \{|U_n(l,k)| > \delta m \} \nonumber \\
& \leq &
\label{inq1}
2n \, \exp \left\{ \frac{-2 \delta^2 m^2}{nL^2(2k+1)^2}
  + \frac{4 \delta m}{n L(2k+1)} \right\} \\[.08in]
& \leq &
\label{inq2}
2n \, \exp \left\{ \frac{-\delta^2 m^2}{2nL^2(2k+1)^2} \right\}
\mbox{ if $\delta m \geq 2L(2k+1)$}. \ \ \ \
\end{eqnarray}

\vskip.05in

\noindent
Inequality (\ref{inq2}) is used below, in conjunction
with the Borel Cantelli Lemma, to establish the almost sure
consistency of the estimates $\hat{x}_{l,n}$ and $\tilde{x}_{l,n}$.
Neither result makes full use of the inequality, which shows,
for example, that for each $\alpha \in (0,1/2)$,
\be
\label{rate}
n^{\alpha} \cdot
\max_{l \in L_n(m,k)}
\frac{|\sum_{j \in A_n(l,k)} \varepsilon_j|}{|A_n(l,k)|}
\ \to \ 0
\ee
with probability one,
provided that $k = O(\log n)$ and
$m \geq n^\beta$ with $\beta \in (\alpha + 1/2,1)$.
The next lemma appears in (\cite{Lall99}); we include the
proof for completeness.

\begin{lem}
\label{comb}
If $k = o(\log n)$ then for every $\epsilon > 0$,
\[
\frac{1}{n} \sum_{j=0}^{n} I
\{ |A_n(j,k)| \leq n^{1 - \epsilon} \} \to 0
\mbox{ as } n \to \infty.
\]
\end{lem}

\noindent
{\bf Proof:} As $\Lambda$ is compact, there exists a finite
set set $S \subseteq \Lambda$, such that
\[
\max_{u \in \Lambda} \min_{v \in S} |u - v|
\leq
\frac{\triangle}{10} .
\]
Let $S^{2k+1}$ be the collection of sequences
$\underline{s} = (s_{-k},\cdots,s_{k} )$ with $s_i \in S$.
For each $x \in \Lambda$ there is
some $\underline{s} \in S^{2k+1}$, such that
$\max_{|i| \leq k} |s_i - F^i x| < \triangle / 10$.
Thus if we define
\[
J_n(\underline{s})
\ = \
\left\{ j : 0 \leq j \leq n \ \mbox{ and } \
\max_{|i| \le k} |s_i - F^{i+j} x|
\leq \frac{\triangle}{10} \right\}
\ \ \ \ \underline{s} \in S^{2k+1}
\]
then each integer $j = k,\ldots, n-k$ is in contained
in at least one set $J_n(\underline{s})$.  Moreover,
if $j_1, j_2 \in J_n(\underline{s})$ then
\[
\max_{|i| \leq k} |x_{j_1+i} - x_{j_2+i} |
< \frac{\triangle}{5}
\ \ \mbox{ and } \ \
\max_{|i| \leq k} |y_{j_1+i} - y_{j_2+i} |
< \frac{3\triangle}{5} ,
\]
and therefore $j_1 \in A_n(j_2, k)$ and $j_2 \in A_n(j_1, k)$.
It follows from this last observation that
$|A_n(j,k)| \leq N$ and $j \in J_n(\underline{s})$ imply
$|J_n(\underline{s})| \leq N$.
Fix $ 0 < \epsilon < 1$.  As $k = o(\log n)$,
$|S^{2k+1}| = |S|^{2k+1} = o(n^{\epsilon/2})$.
Let $\sum_{\underline{s}}$ denote the sum over
$S^{2k+1}$.  When $n$ sufficiently large,
\begin{eqnarray*}
\sum_{j=0}^n I\{ |A_n(j,k)| \leq n^{1-\epsilon} \}
& \leq &
\sum_{j=0}^n \sum_{\underline{s}}
I\{ |A_n(j,k)| \leq n^{1-\epsilon} \}
I\{ j \in J_n(\underline{s}) \} \\
& \leq &
\sum_{\underline{s}} \sum_{j=0}^n
I\{ |J_n(\underline{s})| \leq n^{1-\epsilon} \}
I\{ j \in J_n(\underline{s}) \} \\
& = &
\sum_{\underline{s}}
|J_n(\underline{s})| \,
I\{ |J_n(\underline{s})| \leq n^{1-\epsilon} \} \\
& \leq &
\sum_{\underline{s}} |J_n(\underline{s})|
I \left\{ |J_n(\underline{v})| \leq
n^{1-\epsilon/2} \cdot |S|^{-2k-1} \right\} \\
& \leq &
n^{1-\frac{\epsilon}{2}} .
\end{eqnarray*}
As the last term above is $o(n)$ the
result follows.

\vskip.2in

\noindent
{\bf Proof of Theorem \ref{FXWN}:}
Fix $\beta \in (1/2,1)$.  The stochastic term in inequality
(\ref{averbnd}) can be bounded as follows:
\[
\frac{1}{n-2k} \sum_{i=k}^{n-k} \,
\frac{|\sum_{j \in A_n(l,k)} \varepsilon_j \, |}{|A_n(l,k)|}
\ \leq \
\max_{l \in L_n(n^\beta,k)}
\frac{|\sum_{j \in A_n(l,k)} \varepsilon_j \, |}{|A_n(l,k)|}
\ + \
\frac{\Delta}{5 (n-2k)} \sum_{i=k}^{n-k} \,
I\{ |A_n(i,k)| \leq n^\beta \}
\]
Inequality (\ref{inq2}) ensures that the first term
on the right hand side tends to zero with probability one.
The second term tends to zero by Lemma \ref{comb}.

\begin{prop}
\label{mww}
Let the window widths $k_l$ be defined as in (\ref{aww}).
If $x$ is strongly recurrent then
$\min\{ k_{j,n} : \, \log n \leq j \leq n - \log n \} \to \infty$
as $n \to \infty$.
\end{prop}

\noindent
{\bf Proof:}
Let ${\cal O}$ be a given finite cover of $\Lambda$ by sets
having diameter less than $\Delta / 5$.
Fix $K \geq 1$ and define $\gamma$ to be the set of all Cartesian
products $O_{-k} \times \cdots \times O_k$ with
$1 \leq k \leq K$ and such that each $O_i \in {\cal O}$.
Let $C_k$ denote
any product of $2k+1$ sets from ${\cal O}$.
As $x$ is assumed
to be strongly recurrent,
$\gamma = \gamma_0 \cup \gamma_1$ where
\[
\gamma_0 \, = \, \bigcup_{k=1}^K \left\{ C_k : \sum_{i=k}^\infty
              I\{ x_{i-k}^{i+k} \in C_k \} < \infty \right\}
\ \ \
\gamma_1 \, = \,
\bigcup_{k=1}^K \left\{ C_k : \liminf_{n} \frac{1}{n}
             \sum_{i=k}^n I\{ x_{i-k}^{i+k} \in C_k \} > 0 \right\}
\]
As $\gamma_0$ is finite, there exists an integer $N < \infty$ such that
$x_{j-k}^{j+k} \in C_k \in \gamma_1$
for every $k \leq K$ and every $j \geq \log N$.  Moreover,
if $x_{i-k}^{i+k}$ and $x_{j-k}^{j+k}$ lie in the same set
$C_k \in \gamma_1$ and $\log n \leq i,j \leq n - \log n$ then
it is clear that $i \in A_n(j,k)$.  Thus when $n \geq N$,
\[
|A_n(j,k)|
\ \geq \
\min_{C_k \in \gamma_1} \sum_{i=k}^{n-k} I\{ x_{i-k}^{i+k} \in C_k \}
\ \ \mbox{ for } \ \ k \leq K \mbox{ and } j = \log n, \ldots, n - \log n .
\]
The definition of $\gamma_1$ ensures that $|A_n(j,k)| \geq n / \log n$
for $n$ sufficiently large and $k,j$ as above.  Therefore
$\liminf_n \min_j k_{j,n} \geq K$ and the result follows as
$K$ was arbitrary.

\vskip.2in

\noindent
{\bf Proof of Theorem \ref{VWIN}:}
Let $\kappa = \log n$ and $m = n / \log n$.  It follows from
inequality (\ref{xhinq1}) and the definition of $k_{l,n}$
that
\begin{eqnarray*}
\max_{\kappa \leq l \leq n-\kappa} | x_l - \hat{x}_{l,n} |
& \leq &
\max_{\kappa \leq l \leq n-\kappa} H^{-1}(k_{l,n}) \ + \
\max_{\kappa \leq l \leq n-\kappa}
\frac{|\sum_{j \in A_n(l,k_{l,n})} \varepsilon_j \, |}
          {|A_n(l,k_{l,n})|} \\
& \leq &
H^{-1} \left(\min_{\kappa \leq l \leq n-\kappa} k_{l,n} \right) \ + \
\max_{1 \leq k \leq \kappa} \, \max_{l \in L_n(m,k)}
\frac{|\sum_{j \in A_n(l,k)} \varepsilon_j \, |}{|A_n(l,k)|}
\end{eqnarray*}
If $x$ is strongly recurrent then the first term on the right hand
side tends to zero by an application of Proposition \ref{FST}
and Proposition \ref{mww}.  Inequality (\ref{inq2}) and
a standard Borel-Cantelli argument show that the second term
tends to zero with probability one.

\section{Proof of Theorem \protect{\ref{HOM}}}
\label{PFH}

Throughout this section $(x,x')$ is a fixed strongly homoclinic
pair for $F$.  Define $x_i = F^i x$, $x_i' = F^i x'$,
$y_i = x_i + \varepsilon_i$ and $y_i' = x_i' + \varepsilon_i$
as above.
As $(x,x')$ is
strongly homoclinic,
\be
\label{sum}
\sum_{i \in \integ} |x_i - x_i'| \ < \ \infty .
\ee

\begin{lem}
\label{acset}
If conditions (\ref{cond2}) and (\ref{cond3}) hold,
then there exist sets $A_i^* \subseteq \real^d$, $i \in \integ$,
such that
\begin{enumerate}

\item[a.] $A_i^* \subseteq (S + x_i) \cap (S + x_i')$ for each $i$, and

\item[b.] $P\{ y_i \in A_i^* \ \mbox{and} \ y_i' \in A_i^*
\ \mbox{for all} \ i \in \integ \} \ > \ 0$.

\end{enumerate}
\end{lem}

\noindent
{\bf Proof:}
For each $i \in \integ$ define $A_i = (S + x_i) \cap (S + x_i')$.
Note that
\begin{eqnarray*}
P\{ y_i \not\in A_i \ \mbox{ or } \ y_i' \not\in A_i \}
& \leq &
P\{ y_i \not\in A_i \} + P\{ y_i' \not\in A_i \} \\
& = &
P\{ \varepsilon_i \not\in (S + (x_i' - x_i)) \} +
P\{ \varepsilon_i \not\in (S + (x_i - x_i')) \} \\
& = &
\eta(\, S \setminus (S + (x_i' - x_i)) \, ) +
\eta(\, S \setminus (S + (x_i - x_i')) \, )  \\
& \leq &
2 \, \eta(\, (\partial S)^{|x_i-x_i'|} \, )
\end{eqnarray*}
Assumption (\ref{cond2}) implies that
$\eta(\, (\partial S)^{|x_i-x_i'|} \, ) \leq c \, |x_i - x_i'|$
for some constant $c < \infty$, and it then follows
from (\ref{sum}) that
\[
\sum_{i \in \integ}
P\{ y_i \not\in A_i \ \mbox{ or } \
    y_i' \not\in A_i \} \ < \ \infty.
\]
By an application of the Borel Cantelli Lemma,
there exists an integer $N$ such that
\be
\label{bcl}
P\{ y_i \in A_i \ \mbox{and} \ y_i' \in A_i
\ \mbox{ for all } \ |i| > N \} \ \geq \ 1/2.
\ee
Define $A_i^* = A_i$ for $|i| > N$.  Clearly
(a) holds for each $|i| > N$.

It remains to select sets $A_i^*$ for $|i| \leq N$.  To this end,
let $v^* \in \real^d$ be any vector such that
for some $\delta > 0$
\[
\sup_{v \in \Lambda} |v - v^*| \ \leq \ \rho - \delta
\]
and define $A_i^* = B(v^*, \, (\rho + \delta)/2 )$ for $|i| \leq N$.
Then for each $v \in \Lambda$,
\[
\sup_{u \in (A_i^* - v)} |u|
\ \leq \
\frac{\rho + \delta}{2} \, +  \, |v^* - v|
\ < \
\frac{3}{2} \, \rho ,
\]
which implies that
$(A_i^* - v) \, \subseteq \, B(0, 3\rho/2) \, \subseteq \, S$.
Thus (a) holds for $|i| \leq N$.  Moreover, for each such $i$,
\begin{eqnarray*}
P\{ y_i \in A_i^* \ \mbox{and} \ y_i' \in A_i^* \}
& = &
P\{ \varepsilon_i \in (A_i^* - x_i) \cap (A_i^* - x_i')  \} \\
& = &
\eta((A_i^* - x_i) \cap (A_i^* - x_i')) .
\end{eqnarray*}
The inequality
$| \, |v^* - x_i| - |v^* - x_i' | \, | \leq |x_i - x_i'| \leq \rho$
implies that $(A_i^* - u_i) \cap (A_i^* - v_i)$
has positive Lebesgue measure.  As the intersection is also
contained in $S$,
the last probability above is greater than zero.
Conclusion (b) of the lemma follows from this observation
and (\ref{bcl}), as the $\varepsilon_i$'s are independent.

\vskip.3in

Let $Q$ and $Q'$ be probability measures on $(\X, {\cal S})$
equal to the respective probability distributions of
the random elements $\by$ and $\by'$.
Using the sets $A_i^*$ from Lemma \ref{acset}, define the
Cartesian product
\be
\label{lamdef}
\Gamma \ = \ \prod_{i \in \integ} A_i^* \ \in \ {\cal S} .
\ee
It follows from part (b) of Lemma \ref{acset} that
$Q(\Gamma), Q'(\Gamma) > 0$.

\begin{lem}
\label{macl}
The measures $Q$ and $Q'$ are mutually absolutely continuous on
$\Gamma$: for each $B \in {\cal S}$ contained in $\Gamma$,
$Q(B) = 0$ if and only if $Q'(B) = 0$.
\end{lem}

\noindent
{\bf Proof:}
Let ${\cal S}_n \subseteq {\cal S}$ denote the sigma field
generated by the coordinate functions $\pi_i(\bx) = x_i$,
with $|i| \leq n$.
Let $Q_n$ and $Q_n'$ be the restrictions of $Q$ and $Q'$
to ${\cal S}_n$, respectively.  Then clearly
\[
dQ_n(\bv) \, = \prod_{i=-n}^n f(v_i - x_i) \, dv_{-n} \cdots dv_n
\ \mbox{ and } \
dQ_n'(\bv) \, = \prod_{i=-n}^n f(v_i - x_i') \, dv_{-n} \cdots dv_n .
\]
Furthermore, Lemma \ref{acset} ensures that
$Q_n$ and $Q_n'$ are mutually absolutely continuous on
$\Gamma$, with derivative
\[
\frac{dQ_n}{dQ_n'}(\bv) \ = \
\prod_{i=-n}^n \frac{f(v_i - x_i)}{f(v_i - x_i')}
\hskip.3in \bv \in \Gamma .
\]
For each $n \geq 1$ let
$\Gamma_n = \{ \bv : v_i \in A_i^* \mbox{ for } |i| \leq n \}$,
and define the ${\cal S}_n$-measurable function
\[
R_n(\bv) \ = \ \frac{dQ_n}{dQ_n'}(\bv) \cdot I\{ \bv \in \Gamma_n \} .
\]
Suppose that $B \in {\cal S}_n$.  Then clearly
$B \cap \Gamma_{n+1} \in {\cal S}_{n+1}$
and $B \cap \Gamma_n \in {\cal S}_n$, and therefore
\begin{eqnarray*}
\int_B R_{n+1} \, dQ'
& = &
\int_{B \cap \Gamma_{n+1}} \frac{dQ_{n+1}}{dQ_{n+1}'} \ dQ'
\ = \
Q_{n+1}(B \cap \Gamma_{n+1}) \\
& = &
Q(B \cap \Gamma_{n+1})
\ \leq \
Q(B \cap \Gamma_n)
\ = \
\int_B R_n \, dQ' .
\end{eqnarray*}
Thus $(R_n,{\cal S}_n)$ is a non-negative super-martingale.
By the martingale convergence theorem, $R_n$ converges with $Q'$-probability
one to a non-negative random variable $R^*$.

We now wish to establish the following relation, which will
imply that $Q' << Q$ on $\Gamma$ (see the argument below):
\be
\label{Qprime0}
Q'\{ \bv \in \Gamma : R^*(\bv) = 0 \} = 0 .
\ee
By condition (\ref{cond1}) there exists numbers $\delta_0 > 0$ and
$c < \infty$ such that
\be
\label{cond1'}
\int_{S \cap (S-z)} \left| \log \frac{f(w+z)}{f(w)} \right| f(w) \, dw
\ \leq \ c \, |z|
\ee
whenever $|z| \leq \delta_0$.  By (\ref{sum}) there is
an integer $m$ such that
$|u_i - v_i| \leq \delta_0$ for $|i| \geq m$.
As $R_m(\bv) > 0$ for each $\bv \in \Gamma$, the equality
(\ref{Qprime0}) will follow from
\be
\label{Qprime1}
\int_\Gamma \left| \log \frac{R^*}{R_m} \right| dQ' \ < \ \infty .
\ee
To establish (\ref{Qprime1}), note that by Fatou's lemma
\begin{eqnarray*}
\int_\Gamma \left| \log \frac{R^*}{R_m} \right| dQ'
& = &
\int_\Gamma \liminf_{n \to \infty}
             \left| \log \frac{R_n}{R_m} \right| dQ'
\ \, \leq \ \,
\liminf_{n \to \infty} \int_\Gamma
\left| \log \frac{R_n}{R_m} \right| dQ'  \\ [.1in]
& \leq &
\liminf_{n \to \infty} \int_{\Gamma_n}
\left| \log \frac{R_n}{R_m} \right| dQ'
\ \, = \ \,
\liminf_{n \to \infty}
\int_{\Gamma_n} \left| \log \frac{R_n}{R_m} \right| dQ_n' .
\end{eqnarray*}
Moreover, for each $n > m$,
\begin{eqnarray*}
\int_{\Gamma_n} \left| \log \frac{R_n}{R_m} \right| dQ_n'
& = &
\int_{\Gamma_n}
\left| \sum_{m \leq |i| \leq n}
\log \frac{f(v_i - x_i)}{f(v_i - x_i')} \, \right|
\prod_{j=-n}^n f(v_j - x_j') \, dv_{-n} \cdots dv_n \\
& \leq &
\sum_{m \leq |i| \leq n}
\int_{A_i^*} \left| \, \log \frac{f(v_i - x_i)}{f(v_i - x_i')} \, \right|
f(v_i - x_i') \, dv_i .
\end{eqnarray*}
By an elementary change of variables, our choice of $m$ and the
inequality (\ref{cond1'}) imply that
\[
\int_{A_i^*} \left| \, \log \frac{f(v_i - x_i)}{f(v_i - x_i')} \, \right|
f(v_i - x_i') \, dv_{i}
\ \leq \ c \, |x_i - x_i'| .
\]
Combining the results of the last three displays, it follows that
\[
\int_\Gamma \left| \log \frac{R^*}{R_m} \right| dQ'
\ \leq \ \sum_{i= -\infty}^{\infty} |x_i - x_i'| .
\]
The sum is finite by (\ref{sum}), which establishes (\ref{Qprime1})
and the relation (\ref{Qprime0}).

Suppose now that $B \in {\cal S}$ is such that
$B \subseteq \Gamma$ and $Q'(B) > 0$.
For $n \geq 1$ define events
$B_n = \{ \bv : \exists \bv' \in B \mbox{ s.t.\ } v_i = v_i'
                \mbox{ for } |i| \leq n \} \supseteq B$.
By another application of Fatou's Lemma,
\begin{eqnarray*}
Q(B)
& = &
\lim_{n \to \infty} Q(B_n) \ = \ \liminf_{n \to \infty} Q_n(B_n)
\ = \
\liminf_{n \to \infty} \int \frac{dQ_n}{dQ_n'} I_{B_n} dQ' \\ [.06cm]
& \geq &
\int \liminf_{n \to \infty} \frac{dQ_n}{dQ_n'} I_{B_n} dQ'
\ \geq \
\int_B R^* \, dQ' .
\end{eqnarray*}
The last inequality above follows from the definition of $R^*$ and
the fact that $B_n \supseteq B$.
As $Q'(B) > 0$, the relation (\ref{Qprime0}) implies that the last
integral and $Q(B)$ are positive.  Thus
$Q' \ll Q$ on $\Gamma$.  An identical argument, exchanging the
roles of $Q$ and $Q'$, shows that $Q \ll Q'$ on $\Gamma$ as well.

\vskip.1in

\begin{lem}
\label{homoclin}
If $(x,x')$ is homoclinic and conditions
(\ref{cond1})--(\ref{cond3}) hold, then for every measurable
function $\phi : \X \to \real^d$,
\be
\label{ebnd}
E[ \, | \phi(\by) - x \, | \, + \,
      | \phi(\by') - x' \, | \, ]
\ \geq \
|x - x'| \, \int_\Gamma \min\left[ \frac{dQ'}{dQ}, 1 \right] dQ
\ > \ 0
\ee
where $\Gamma \subseteq \X$ is defined as in (\ref{lamdef}).
\end{lem}

\noindent
{\bf Proof:}
Lemma \ref{macl} shows that $Q' \ll Q$ on $\Gamma$.  Let
$(dQ'/dQ)(\bv)$ be the associated derivative, which is
defined for each $\bv \in \Gamma$.
The expectation above can be written equivalently as
\begin{eqnarray*}
\int | \phi - x | \, dQ \, + \,
\int | \phi - x' | \, dQ'
& \geq &
\int_\Gamma | \phi - x | \, dQ \, + \,
\int_\Gamma | \phi - x' | \, dQ' \\
& = &
\int_\Gamma \left[
   | \phi - x | \, + \,
   | \phi - x' | \, \frac{dQ'}{dQ} \, \right] dQ \\
& \geq &
|x - x'| \, \int_\Gamma \min\left[ \frac{dQ'}{dQ}, 1 \right] dQ .
\end{eqnarray*}
As $(dQ'/dQ)(\bv)$ is positive for $Q$-almost every
$\bv \in \Gamma$, and $Q(\Gamma) > 0$,
the last integral is positive.

\vskip.2in

The lower bound in Lemma \ref{homoclin}
bears further discussion.  Suppose for
the moment that the distribution $\eta$ of the noise satisfies
(\ref{cond1}) and is supported on all of $\real^d$, which is the case \
if the $\varepsilon_i$ are Gaussian.  Then
we may take $A_i^* = \real^d$ for each $i$,
so that $\Gamma = \X$.  In this case,
further evaluation leads to a
simplification of the integral in (\ref{ebnd}):
\begin{eqnarray*}
\int \min\left[ \frac{dQ'}{dQ}, 1 \right] dQ
& = &
Q\left\{ \frac{dQ'}{dQ} \geq 1 \right\}
\ + \
Q'\left\{ \frac{dQ'}{dQ} < 1 \right\} \\
& = &
1 \ - \
Q\left\{ \frac{dQ'}{dQ} < 1 \right\}
\ + \
Q'\left\{ \frac{dQ'}{dQ} < 1 \right\} \\
& = &
1 \ - \ || Q - Q' || .
\end{eqnarray*}
Here $|| Q - Q' || \ = \ \sup_{B \in {\cal S}} |Q(B) - Q'(B)|$
is the total variation distance between $Q$ and $Q'$.  As $Q$ and
$Q'$ are mutually absolutely continuous, $|| Q - Q' || < 1$ and
we see again that the lower bound in Lemma \ref{homoclin} is positive.
When $\Gamma \neq \X$ one may derive a similar, but more
complicated, expression for the integral in (\ref{ebnd}).

Although no scheme can reliably distinguish between the elements of
a strongly homoclinic pair $(x,x')$ from noisy observations
of their trajectories, we may say that a scheme $\phi$
is optimal for this pair if it achieves the lower bound above.
One may readily check that the maximum likelihood scheme
\[
\phi(\bv) \ = \
\left\{ \begin{array}{ll}
        x & \mbox{ if } \ \frac{dQ'}{dQ}(\bv) \leq 1 \\
        x' & \mbox{ otherwise }
        \end{array} \right.
\]
is optimal in this sense.

\bibliographystyle{plain}

\small{

\end{document}